\pdfoutput=1
\documentclass[sigconf]{acmart}
\usepackage[utf8]{inputenc}
\usepackage{booktabs}
\usepackage{soul}

\usepackage[ruled,vlined]{algorithm2e}
\usepackage{amssymb}
\usepackage[]{braket}
\newlength\mylen
\newcommand\myinput[1]{%
  \settowidth\mylen{\KwIn{}}%
  \setlength\hangindent{\mylen}%
  \hspace*{\mylen}#1\\}

\setcopyright{rightsretained}

\begin{document}
\title{Spatio-Temporal Reference Frames as Geographic Objects}

\author{Andrew Simmons}
\orcid{0000-0001-8402-2853}
\affiliation{%
  \institution{Deakin University}
  \streetaddress{221 Burwood Highway}
  \city{Burwood}
  \state{Victoria}
  \country{Australia}
  \postcode{3125}
}
\email{a.simmons@deakin.edu.au}

\author{Rajesh Vasa}
\affiliation{%
  \institution{Deakin University}
  \streetaddress{221 Burwood Highway}
  \city{Burwood}
  \state{Victoria}
  \country{Australia}
  \postcode{3125}
}
\email{rajesh.vasa@deakin.edu.au}

\renewcommand{\shortauthors}{A. Simmons and R. Vasa}

\begin{abstract}
  It is often desirable to analyse trajectory data in local coordinates relative to a reference location. Similarly, temporal data also needs to be transformed to be relative to an event. Together, temporal and spatial contextualisation permits comparative analysis of similar trajectories taken across multiple reference locations. To the GIS professional, the procedures to establish a reference frame at a location and reproject the data into local coordinates are well known, albeit tedious. However, GIS tools are now often used by subject matter experts who may not have the deep knowledge of coordinate frames and projections required to use these techniques effectively.

  We introduce a novel method for representing spatio-temporal reference frames using ordinary geographic objects available in GIS tools. We argue that our method both reduces the number of manual steps required to reproject data to a local reference frame, in addition to reducing the number of concepts a novice user would need to learn.
\end{abstract}

\copyrightyear{2017}
\acmYear{2017}
\setcopyright{rightsretained}
\acmConference{SIGSPATIAL'17}{November 7--10, 2017}{Los Angeles Area, CA, USA}\acmDOI{10.1145/3139958.3139983}
\acmISBN{978-1-4503-5490-5/17/11}



\maketitle

\section{Introduction}

With the widespread availability of GNSS technologies, modern data sets are often created and stored directly using a global coordinate system such as WGS84. Furthermore, collecting and analysing spatial data is moving beyond surveyors and cartographers to subject matter experts in their own field, from traffic engineers to sport performance analysts. However, in many applications, it is position relative to a frame of reference that matters, and not absolute geographic location. Analysing data for such applications requires the data to be reprojected from the global coordinate system into a local coordinate system. Unfortunately, this task requires deep conceptual knowledge to perform, and may bar novice users from unlocking the full potential of their data.

Consider a sport performance analyst conducting spatial analysis of player GPS tracking data, who wishes to analyse spatial patterns for a team over the course of the sport season. A single game alone does not contain enough events to mine statistically significant patterns. In order to make the most of the team's dataset, the sport analyst will need to consolidate data from each of the sport fields that the team competes at. To achieve this, the sport analyst will need to reproject the raw GPS traces from games at multiple sport fields into a consistent local coordinate system. This requires the sport performance analyst to define a projection for each sport field the team played on.

While it would be possible to achieve this task using a conventional GIS system, it would require the user to manually write a projection string using a format such as Proj4 or Well Known Text (WKT) for every field that the team played on. Furthermore, expertise working with projections and reference frames is needed to ensure this task is performed correctly, as incorrectly projecting data is a common cause of subtle geospatial errors that may undermine the validity of the analysis.

In this paper, we address these issues through introducing a novel method for defining reference frames using geometric line segment objects. We argue that our system is more learnable for novice GIS users, as it reduces the complexity of defining reference frames to the more familiar task of drawing lines. We further show that the number of manual steps required to perform a task using our approach scales well with the the number of reference frames, trajectories, and events of interest, thus indicating that our approach is productive for large problems.

\section{State of the Art}

\begin{figure*}
\includegraphics[width=1.0\textwidth]{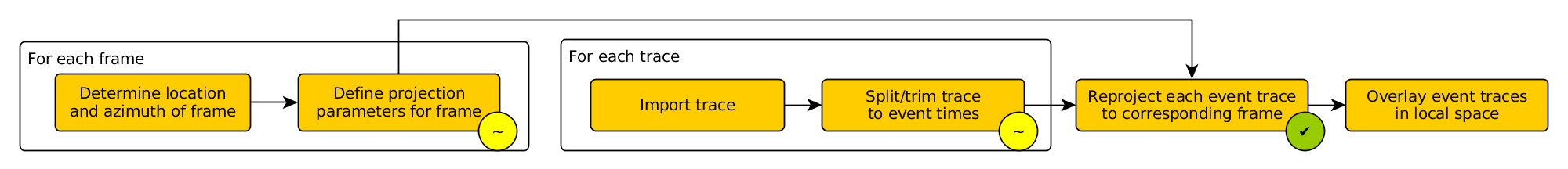}
\caption{Tasks required to split traces (e.g. player GPS trajectories) that span multiple events (e.g. game segments), and reproject relative to frame (e.g. sport field) such that event traces can be compared within a common local coordinate system. Our proposed solution allow tasks marked with a green check to be fully automated, and tasks with a yellow tilde to be semi-automated.}
\label{fig:manual-process}
\end{figure*}

In Fig~\ref{fig:manual-process} we provide the workflow of high level tasks that a user (e.g. a sport performance analyst) would need to complete in order to transform their dataset (e.g. trajectory data for each player of a team) into a common local space so that they fully utilise all event data (e.g. making comparisons of game segments throughout the season collected from multiple sport fields). In this section we will briefly elaborate how tasks would typically be accomplished using existing GIS tools, which will become our baseline for evaluating the merits and drawbacks of our proposed system.

\ul{Determining location and azimuth of frame:} In order to correct for the location and orientation of each frame (e.g. sport field) the user needs to first obtain its location and orientation. For novice users, this seemingly simple step is a common cause of errors, as it is essential to ensure the azimuth is measured using the correct geodesic calculations, rather than simply the angle displayed within the current projection (which is often distorted).

\ul{Define projection parameters for frame:} If frames are separated by a non-negligible distance with respect to the curvature of Earth (e.g. an interstate game), then it is necessary to work with different projections for each region. Attempting to compare geometries across large distances by simply offsetting coordinates can result in distortions as the regions may have different scales within the projection system. While databases of common projection parameters for each segment of the globe exist, achieving minimum distortion requires specifying custom projection parameters. This in turn requires the user to manually write a projection string using a format such as Proj4 or Well Known Text (WKT).

\ul{Split/trim trace to event times:} It is unrealistic to start and end collection of trajectory data for the exact duration of an event of interest. Rather, data is typically logged for a short time before and after the event, or even multiple events (for example, in a sport game, data loggers will be left on for the entire game, and the data can be split into individual round durations afterwards, discarding data for rest breaks between rounds). Manually splitting and trimming this data to the precise event durations can be a tedious process, as well as a repetitive task when there are many traces (e.g. analysing game data for every member of a sport team).

\ul{Reproject each event trace to corresponding frame:} The user needs to manually match up each event trace with the relevant projection parameters to reproject it.

\ul{Overlay event traces in local space:} For this step, the user needs to take the unconventional step of discarding any projection metadata so that event traces can be reinterpreted as local x,y data in a relative coordinate system.

\section{Related Work}

Hägerstrand visualises time as a dimension in order to conceptually reason about constraints on human mobility in space-time \cite{hagerstrand_what_1970} (although it should be noted that the idea of time as a dimension was present as early as the 18th century in the works of d'Alembert and Lagrange \cite{archibald_time_1914, goenner_history_2008}). In order to permit the visualisation to be easily perceived, Hägerstrand uses a 2 dimensional map, thus freeing up the third dimension to visualise the passage of time. The three axes form a space-time cube within which mobility can be analysed.

While Hägerstrand originally presented the space-time cube as a conceptual tool rather than as a practical means of visualizing spatio-temporal datasets, it has inspired a number of novel visualization systems for viewing spatio-temporal data within the time-space cube \cite{kraak_m._space-time_2003, gatalsky_interactive_2004, andrienko_visualization_2014}.

Jenny and Hurni \cite{jenny_studying_2011} describe the series of cartographic transformations needed to setup a common coordinate system to compare historic maps against modern maps. Jenny and Hurni stress the importance of first projecting the modern map into the coordinate system of the old map to prevent subtle distortion artefacts that would arise from comparing data in different projections.

Šavrič et al. created Projection Wizard \cite{savric_projection_2016} to aid novice users with the task of selecting a projection. Their system allows the user to interactively drag a box around the region of interest, and the system will automatically suggest a projection based upon the guidelines suggested by John P. Snyder in his classic work \textit{Map projections: A working manual} \cite{snyder_map_1987}. Unfortunately, their system doesn't allow the user to draw a rotated region, thus limiting its ability to suggest oblique projections.

\section{Proposed Solution}

\begin{figure}
\includegraphics[width=0.5\textwidth]{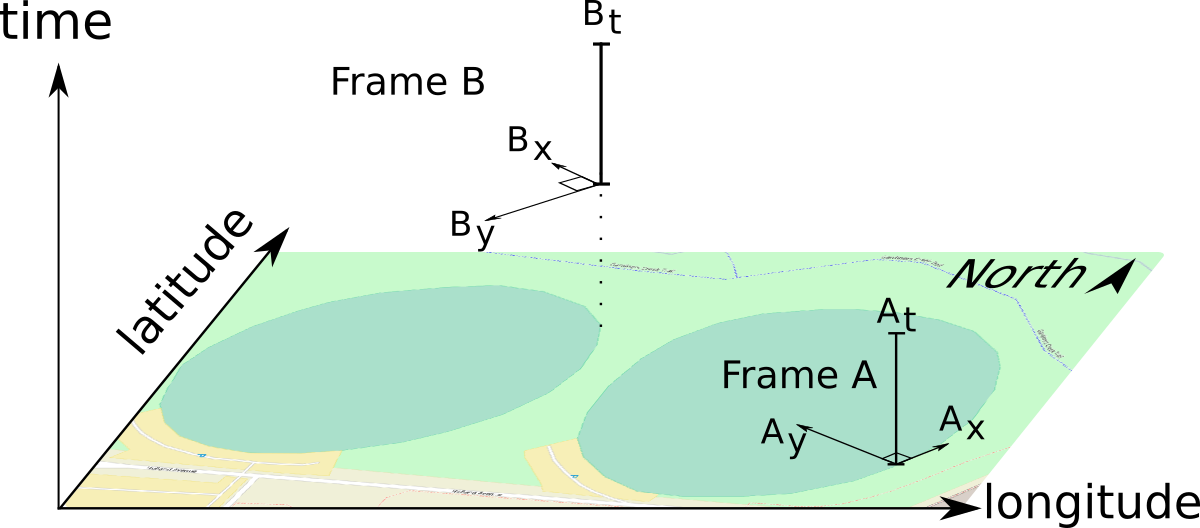}
\caption{Configuration of spatio-temporal reference frames used to describe spatio-temporal structure of a sport game.}
\label{fig:space-ref}
\end{figure}

In this paper, we utilize Hägerstrand's space-time as a conceptual framework for defining reference frames. Our reference frames are orientated in space, as well as positioned in time. Our reference frames serve as objects that users can manipulate to describe the structure of space-time to the computer. This in turn, benefits the user by allowing transformations between global and local coordinates to be automated.

We provide an example of setting up the structure of a sport game in Fig~\ref{fig:space-ref}. In this example, there is an initial game event (e.g. a practice game) played in the rightmost field, and this is later followed by a game event in the leftmost field. A reference frame is added to each sport field, and oriented towards the target goal. This make distances and movements comparable across the two fields. The time dimension of a reference frame spans an event interval (e.g. the start and end time of the round), which will serve as the basis for relative time within that frame. A more complex game would have additional frames for each round. If players change sides of the field after each round, then this could be accounted for by a series of reference frames in time that alternate between each side of the field so as to ensure that the team of interest always appears to be heading the same local direction from the perspective of the reference frame.

\subsection{Representation}

In conventional GIS systems, reference frames are not treated as first-class citizens like ordinary geographic objects such as polygons. Transformations in GIS systems are performed imperatively rather than declaratively: that is, they require the user to manually enter the coordinates and apply the transformation rather than providing a means to symbolically represent the transformation within the GIS system (although GIS systems do allow specifications of procedures using scripts that can be re-run later, this requires programming skills that would be additional expectation of a novice user).

For our system, in addition to the trajectory dataset, we require a spatial file that contains geometries to describe the reference frames, with the temporal aspect of the reference frames represented using property fields of the geometry. Specifically, the spatial aspect of a spatio-temporal reference frame is represented by a line with two points. Our choice to represent spatio-temporal reference frames as ordinary geographic objects allows the user to create and modify the reference frames in the same manner as other data. Thus novice users only need to learn the basic GIS editing features, and can reuse this knowledge to create reference frames in our system without having to learn a new interface specifically for creating reference frames. Furthermore, it means that our system does not require any proprietary modifications to existing GIS tools or GIS data formats beyond the geometry property fields offered by spatial formats such as GeoJSON.

The temporal aspect of a spatio-temporal reference frame is represented by an ISO 8601:2004 time interval declared as an event duration attribute associated with the frame geometry.

 Conceptually, we consider each spatio-temporal reference frame as distinct, consisting of a line (representing the spatial orientation of the frame) and one attribute (representing the temporal aspect of the frame). However, in complex scenarios, there can be many spatio-temporal reference frames with the same spatial dimension (e.g. a game that has multiple rounds occurring at the same location), thus for convenience, we consider a line with multiple event duration attributes to represent a set of spatio-temporal reference frames for each event.

\section{Implementation}

\begin{algorithm}
  \DontPrintSemicolon
  \KwIn{Set of GPS traces for processing, $G$.}
  \myinput{Set of line segment geometries, $L$, representing}
  \myinput{spatial frames.}
  \myinput{Event metadata for each geometry, $Metadata : L \to E$.}
  \KwResult{Projection $Z : G, L, E \to [\mathbb{R}^2]$ of each non-empty permutation of GPS-trace, spatial-frame, event-interval to local $x,y$ coordinates at offset time $t$.}
  \ForEach{trace $g$ in G}{
    $P \leftarrow \set{(p_\phi, p_\lambda, p_t) \mid p \in g}$\;
    \ForEach{line segment $l = ((\phi_1, \lambda_2), (\phi_2, \lambda_2))$ in $L$}{
      $\alpha \leftarrow$ azimuth from $(\phi_1, \lambda_2)$ toward $(\phi_2, \lambda_2)$\;
      \ForEach{event $e = [e_{begin}, e_{end}]$ in $Metadata(l)$}{
        $P' \leftarrow \set{p \in P \mid e_{begin} \leqslant p_t \leqslant e_{end}}$\;
        \If{$P' = \{\}$}{
          continue to the next event\;
        }
        \ForEach{point $p$ in $P'$}{
           $x, y \leftarrow$ project $(p_\phi, p_\lambda)$ to oblique Mercator projection centered at $(\phi_1, \lambda_2)$ with centerline at azimuth $\alpha$\;
           $t \leftarrow p_t - t_{begin}$\;
           $Z(g, l, e)[t] \leftarrow (x, y)$\;
        }
      }
    }
  }
  \caption{GPS to XYT}
  \label{alg:gps2xyt}
\end{algorithm}

\begin{figure*}
\includegraphics[width=0.9\textwidth]{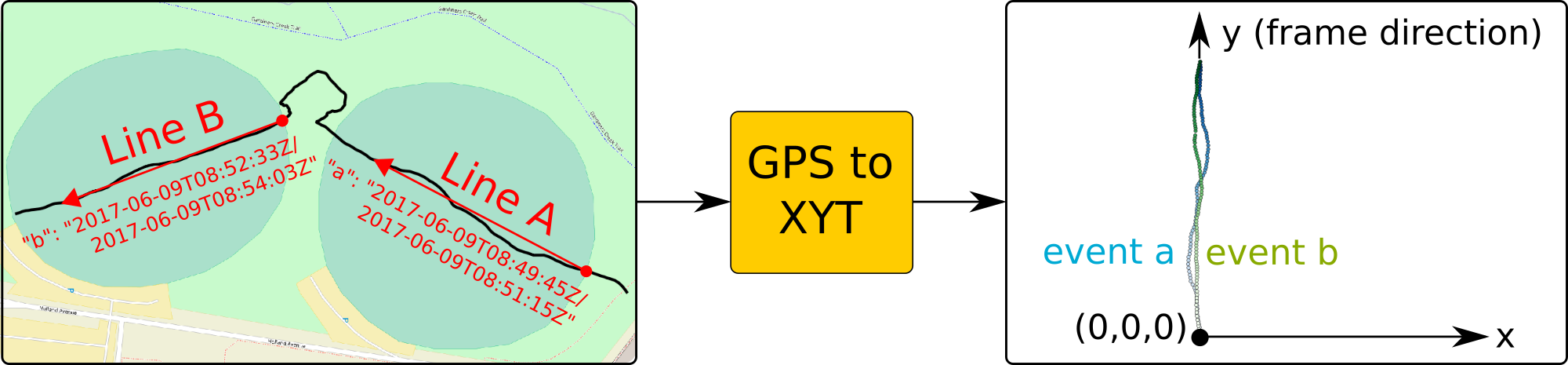}
\caption{(left) Spatio-temporal reference frames are represented as the combination of a line (shown in red, direction is important) and a property attribute to represent temporal interval (small red text). These reference frames are stored in the GeoJSON format and can be created using basic editing features within a GIS tool. The user provides a directory of trajectories (black lines) in the GPX format. Our proof-of-concept \textit{GPS to XYT} (middle box) iterates over each permutation of temporal frame and trajectory, trimming the trajectory to the duration of the event. If any data is found for that duration, it reprojects the trace relative to the corresponding reference frame the event is associated with (see Alg~\ref{alg:gps2xyt} for details). (right) The reprojected data contains relative x,y,time points suitable for comparative analysis in an external system.}
\label{fig:automated-process}
\end{figure*}

We have created a proof-of-concept\footnote{Our code is made publicly available at \texttt{https://github.com/anjsimmo/gps\_xyt}} implementation of our proposed system that given a GeoJSON file containing the spatio-temporal reference frames, and a directory of raw GPS traces (in GPX format) will automate the splitting, reprojection and temporal shifting of event traces into local coordinates. The resultant output is a directory of reprojected events, each containing x,y,t fields that represent relative location perpendicular to the reference frame, relative location in the direction of the reference frame, and relative offset time from the start of the reference event. We present this pipeline graphically in Fig~\ref{fig:automated-process}. The core of this program is a nested loop over each trace, line segment, and event, as shown in Alg~\ref{alg:gps2xyt}.

We utilise the Hotine Oblique Mercator Projection (also known as Rectified Skewed Orthographic) \citep[p. 66]{snyder_map_1987} to perform the projection from global coordinates into a local x,y coordinates relative to the frame. The projection parameters are chosen to (approximately) preserve scale along a centerline passing through the reference frame origin and oriented at the same azimuth as the direction of the reference frame. This means that the reference frame conversions will be reasonable even for frames that cover very long distances in the direction of the reference frame (e.g. the trajectory of a point-to-point air trip). The actual projection calculations are carried out using the Open Source Proj.4 library \cite{evenden_libproj4_2005}.

\section{Evaluation}

\subsection{Bloom's Revised Taxonomy of Learning}

In this section we repurpose Bloom's revised taxonomy \cite{anderson_taxonomy_2001} (intended for evaluating education curricula) as a means of reasoning about the learnability of our system for novice users. Bloom's revised taxonomy acknowledges two dimensions to the complexity of a learning task. The first is the \textit{Cognitive Process Dimension}: `Remember' (low order thinking), `Understand', `Apply', `Analyze', `Evaluate', `Create' (high order thinking). The second dimension is the \textit{Knowledge Dimension}: `Factual' (concrete), `Conceptual', `Procedural', `Metacognitive' (most abstract).

GIS tools and libraries have automated the procedural aspects of projection (Apply + Procedural). However, they still require a deep conceptual knowledge of projection and coordinate systems in order to create projection frames (Create + Conceptual). Our system reduces projection from an abstract concept to geometric objects on the map, represented the same way as other concrete data that the user creates (Create + Factual). This lowering of the knowledge dimension represents a decrease in complexity and less depth of required learning.

\subsection{Productivity}

In addition to simplifying many of the steps articulated in Fig~\ref{fig:manual-process}, our system significantly reduces the number of steps as the number of traces scales up. Manually splitting the trajectory data would require $\mathcal{O}(F \times E \times G)$ steps where $F$ is the number of reference frames, $E$ is the number of event durations of interest, and $G$ is the number of GPS trajectory traces -- for a very large number of traces, the theoretical limit results not from the definition of the reference frames themselves (as these can be reused) but rather from the manual step of reprojecting each combination of frame, trace, and event. Our system still requires the user to provide some information to specify the spatial frames (in the form of a line geometry), and events (in the form of properties attached to the frames). However, our system entirely automates nearly all interaction with GPS trajectory traces (other than placing the trajectory traces in a folder for processing). Thus our system reduces the number of steps to $\mathcal{O}(F \times E + G)$.

\section{Conclusions}

This system solves the fundamental problem of enabling non GIS experts to define and make comparisons between local reference frames, and does so in a general way. Thus our system is applicable to a diversity of domains, e.g. to enable traffic engineers to rapidly compare local traffic conditions taken from floating car data collected at two or more identical housing estates.

Rather than treat reference frames as just an abstract mathematical concept, we have shown that they can be represented as geographic objects within GIS tools. This technique increases the power of GIS, as rather than limiting ourselves to viewing and manipulating low level facts, we can deal with the projection systems themselves as if they were ordinary geographic objects. We have demonstrated that it is trivial to implement a program to bulk reproject data using these user constructed reference frames as input. However, with additional software engineering effort, the algorithm could be used to interactively reproject streams of data or to provide the user with immediate visual feedback on how the final reprojected data will look as they manipulate the reference frame objects. Future work is needed to investigate whether more complex geometries could be used to automatically reproject GPS data with respect to a more abstract projection (such as a schematic transit map), and whether it is possible to automatically extract reference frame objects using landmarks identified in satellite imagery.

\bibliographystyle{ACM-Reference-Format}
\bibliography{lit}

\end{document}